# Complexity Leads to Simplicity: A Consensus Layer V Pyramidal Neuron Can Sustain Interpulse-Interval Coding

Chandan Singh[1], William B Levy[1*]
[1] Departments of Neurosurgery and of Psychology, University of Virginia, Charlottesville, VA, USA.
* Corresponding author
Email: wbl@virginia.edu

**Abstract**

In terms of a single neuron's long-distance communication, interpulse intervals (IPIs) are a possible alternative to rate and binary codes. As a proxy for IPI, a neuron's time-to-spike (TTS) can be found in the biophysical and experimental literature. Using the current, consensus layer V pyramidal neuron, the present study examines the feasibility of IPI-coding and examines the noise sources that limit the information rate of such an encoding. In descending order of noise intensity, the noise sources are (i) synaptic variability, (ii) sodium channel shot-noise, followed by (iii) thermal noise with synaptic noise much greater than the sodium channel-noise. More importantly, the biophysical model demonstrates a linear relationship between input intensity and $TTS^{-1}$. This linear observation contradicts the assumption that a neuron should be treated as a passive, electronic circuit (an RC circuit, as in the Stein model). Finally, the biophysical simulations allow the calculation of mutual information, which is about 3.0 bits/spike.

**Author Summary**

In order to obtain a complete understanding of neural computation and communication, it is necessary to understand how one neuron communicates with another. To create a quantitative understanding of such communication, we need to understand the neural code. One possible code is referred to generically as a spike-timing code, or in a more technical way, an interpulse interval code. We use a biophysical model of a neocortical pyramidal neuron consisting of an appropriate morphology and known voltage-activated channels to qualify this neuron's time-to-spike in response to current injections or simulated synaptic inputs. This biologically appropriate and biophysically complex has a simple linear characterization necessary for interpulse interval coding, a characterization not demonstrated by the simpler passive model often used by neurotheoreticians. Thus, we have an example where biophysical complexity at the level of voltage-activated channels leads to input-output simplicity at the level of a neuron's encoding of information.

**Introduction**

This study addresses three contemporary topics aimed at understanding neural computation and neural codes: (1) Is the leaky integrate-and-fire neuron a better model than a neuron using linear additivity to reach threshold? (2) What are the relative contributions of the stochastic processes that limit information flow across a neuron? And (3) what is the bits-per-spike mutual information capability of a neocortical pyramidal cell using a recent biophysical model of this neuron's spike-generation?

McCulloch and Pitts introduce the computational neuron as a deterministic threshold-linear device [1]. Gerstein and Mandelbrot consider a linearly additive neuron in a stochastic setting [2]. However, even in this novel work which assumes an additive neuron, there is the absence of wholehearted support for linearity because of a neuron's known leak-currents. Indeed, although there has been some biophysical modeling that attempts to justify a linear neuron [3-5], much effort has gone into understanding the stochastic, leaky-neuron model and its generation of a interpulse interval via first hitting-time [6-13].

For neurons functioning within cortical brain circuitry, there is support for a neuron the performs linear integration [14-16]. It has been suggested that nonlinear conductances can offset certain sublinear processes [3]. It has been recognized for m ore than a quarter of a century that the non-linearity offered by voltage-controlled channels can take a sublinear neuron and linearize it, at least in terms of loss of synaptic drive. More recently, preliminary biophysical models can compensate for time-dependent leak [17]. As shown below, this linear hypothesis exists over a limited range, for a biophysical model that accurately fits demanding neurophysiological data concerning the shape and initiation site of the action potential [18]. In particular, the recent biophysical model [19] of a non-stochastic neuron demonstrates a certain restricted linearization. In this context, the neuron selected here seems to be the "consensus pyramidal neuron" with several labs using the model consensus morphology [20-23] and further the same VGCs [19,24] with agreement in terms of the variety and placement of voltage-activated conductances. It includes, in addition to a voltage-controlled K-channel, two types of voltage-controlled Na-channels (Nav 1.2 and Nav 1.6) with a particular microscopic distribution, notably a high concentration of the low-threshold Nav 1.6 in the initial segment of the axon (AIS) [25-28].

Here the deterministic model (continuous Na-channel conductance as a function of voltage) is the starting point of the model. Then, after evaluating the inverse relationship between injected current and time-to-spike, a stochastic model subject to Na-channel shot activations is investigated. This neuron is activated in two different ways: with a noise-free current-step and with a random point-process. Considering three noise processes — thermal, Na-channel shots, and synaptic activation shots — synaptic noise seems to be the only quantitatively meaningful noise-source. It is fairly well-established that synaptic noise dominates [29] and that the most dominant source of noise besides electrical noise is channel-noise [30-32]. Interest seems to center on sodium channel noise, as this is believed to be the largest [33].

Previous calculations of mutual information that indirectly evaluate the effect of intrinsic noise use the Hl model [34]. This model assumes a Gaussian input and a Gaussian output. Such an assumption might be appropriate for a visual world in motion, but does not seem appropriate for the problem of transmitting information of scene analysis during visual fixations. For such problems one can propose the existence of a scalar latent variable and a conjectured output code. With these two hypotheses, Shannon's mutual information should consider the transmission of a scalar latent variable that parameterizes a Poisson process as the input while the encoded output of the neuron is interpulse interval. We consider the information throughput of a contemporary biophysical model of a neocortical pyramidal neuron and analyze it in a region that approximates the interpulse interval encoding of a neuron.

Such a noise analysis may or may not be intrinsically interesting, but we contend that the role such noise plays in the bit-rate of a neuron is a matter of interest. Typically, (for another method, see [34]) to analyze the bit-rate via Shannon's mutual information requires a statement of the code (an input and an output variable) and a quantitative understanding how stochastic aspects affect transmitting an input value via this code. At the extreme, it is well understood that thermal noise will always keep mutual information finite. However, more precise statements require much better knowledge of all the relevant stochastic processes and some amount of biophysics. Here, using the biophysical model, the mutual information is estimated to be about 3.0 bits per spike.

## Results
### Linearized inverse time-to-spike

The physiological origin of the Hu et al model seems based on a very narrow band of current-step intensities. In fact, the model sustains an inverse linear range for current-steps ranging from 0.51 nA to 0.85 nA when there is the requirement that action potential initiation begins at the AIS; relaxing this last requirement extends the range ever so slightly (the upper bound becomes 1.0 nA). Fig 1 illustrates this inverse relationship. Fig 1A uses the point-injection of a current-step. In this case the slope is 0.30 $\mu C^{-1}$ with an extrapolated y-intercept at zero-current of -0.13 $ms^{-1}$. When spatially-distributed synaptic activation produces the action potential, the slope is 0.0028 $events^{-1}$ with a zero-intensity intercept of -0.09 $ms^{-1}$. Perhaps due to the none point-nature of synaptic activation, the currents required to fire using synaptic activation are consistently larger than the point current injections. That is, assuming a current-step of 1 nA, dividing by a driving voltage of 50 mV, then dividing by the synaptic conductance of 200 pS, and finally dividing by the synaptic event duration of 1.2 ms, yields 83.3 events / ms.

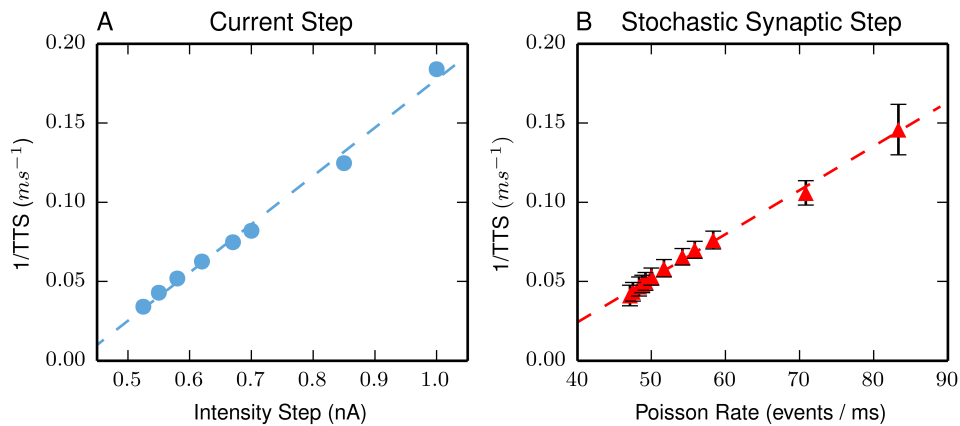

**Fig 1. A linear inverse TTS as a function of excitation.** Excitation is either (A) a point dendritic current-step or (B) a spatially dispersed, synaptic activation. Lines are best linear fits (see text). Each point is an average of 120 excitations from rest. The error bars (SEM) for the current-step are within the plot points. All points but the highest intensities always had spike initiation at the AIS. At the largest intensity on each curve, the spike originated in the dendrite 20 percent of the time.

**Variation in TTS**

The overriding variation of the TTS using synaptic activation arises from the variability of the stochastic process itself. This conclusion is best seen by comparing the histogram of TTS for a deterministic current-step to the stochastic synaptic activation step.

For the highest intensities, the ratio of the variances is more than 200, while for the lowest intensities the ratio of is the variances is more than 40. The standard deviations are plotted in Fig 2 for better visualization of the size of this effect as a function of inverse intensity (note when viewing these curves, the leftmost values of TTS correspond to the highest intensities). The noise increases as the TTS increases; this is mostly because the number of events needed to fire a spike increases with the TTS. In addition, threshold starts to rise. Such a greater amount for depolarization means that more sodium channels need to be activated at threshold. This larger number of sodium channels correlates with a larger standard deviation. Nevertheless, there is no spontaneous firing in this neuron; the contribution of an individual channel fluctuation of sodium is not enough to fire the neuron until the neuron is very close to threshold.

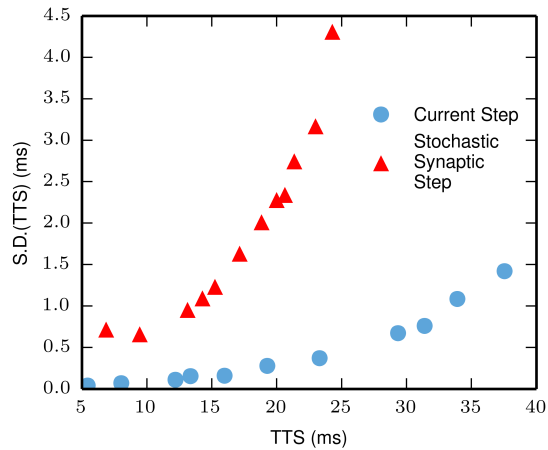

**Fig 2. Synaptic activation increases variation.** The only variation in TTS using a current-step is due to the stochastic nature of Na-channel activation. Random synaptic activation greatly increases the variation in TTS. Plot points correspond to the reverse-ordered successive intensities of Fig 1.

As is well-known for additive point processes, a larger number of events with smaller values (conductances) leads to lower average variance of the steady-state voltage. This lower variance in the voltage is reflected as lower variance in the TTS. A simple model of this relationship between steady-state variance of the voltage and the variance of the TTS (as a first hitting-time distribution) is seen in the standard result [35] that the variances of a first hitting-time and of a Brownian motion (which shot noises can approximate) are proportional.

As one would expect, maintaining a constant total conductance while changing the pS/event changes the variance. In fact, there is a linear relationship between the size of the individual conductance event and the variance (see Fig 3). The best linear fit for Fig 3A is Var[TTS] = 0.0012*x + 0.003 with an $R^2$ value of 0.984 and the fit for Fig 3B is Var[TTS] = 0.0052*x - 0.0797 with an $R^2$ value of 0.993. The conductance of a sodium channel is in the vicinity of 10 – 20 pS [36-38], while the conductance of a synapse is about 200 pS. Synaptic noise dominates, even when varying the parameters of the model (ex. pS/channel, pS/synapse). When combining both sources of noise, the noise is indistinguishable from the noise generated only by the synaptic input.

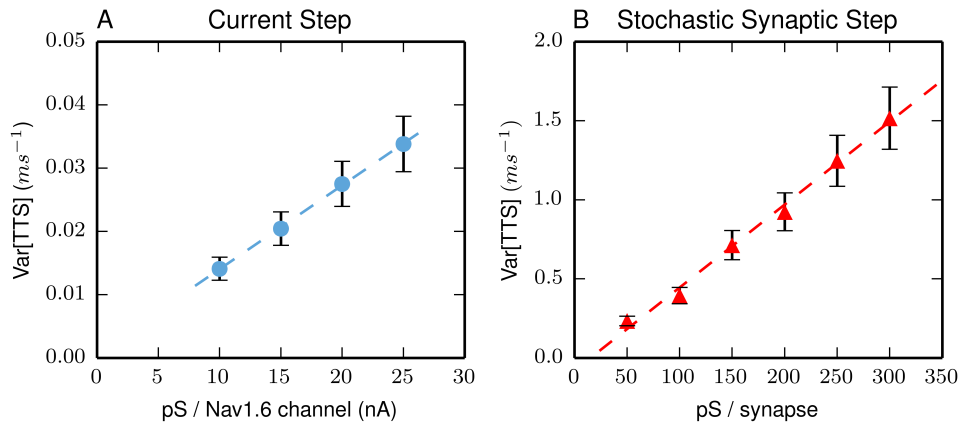

**Fig 3. Synaptic shot-noise far exceeds Na-channel shot-noise** (note the y-axis scale differences). (A) is a current-step of 0.7 nA and (B) has λ=58.3 events/ms for a mean pS/synapse of 200. TTS variance increases as conductance events get larger while keeping $\bar{g}$ constant. Fixed synaptic excitation at 200 pS/synapse results in Var[TTS]=0.92 ms$^2$. Although variance in shot-noise sums should increase as the square root of increasing individual conductances, TTS variance is linear. Error bars are SEM. Lines are best linear fits (see text).

Finally, as will be important in the next section, it is necessary to have reliable probability distributions for the relationship between intensity and TTS. Fig 4 again shows that synaptic variation swamps sodium channel-induced variation (whatever noise is produced in the current-step histogram is incorporated into the stochastic synaptic step). However, more to the point is that the TTS distribution can be fit by an inverse Gaussian distribution (see Fig 4).

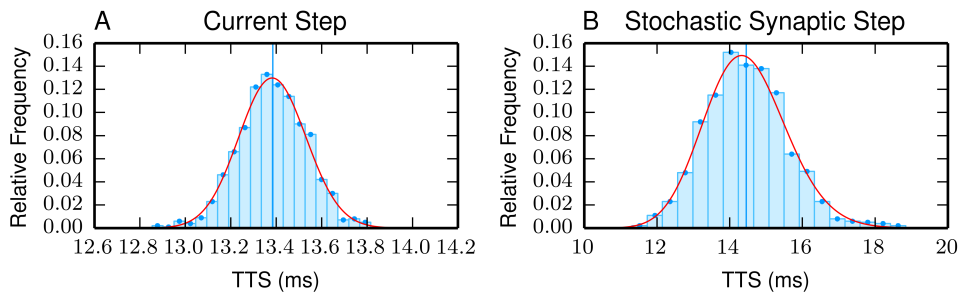

**Fig 4. TTS relative frequency histogram and overlaid inverse Gaussian distribution** with the same mean and variance. (A) Using a current-step of .67 nA, the mean TTS is 13.38 ms (vertical line) and the variance is 0.022 ms$^2$. (B) Using Poisson synaptic activation (λ=55.8 events/ms), the mean TTS is 14.46 ms (vertical line) and the variance is 1.25 ms$^2$. 1,000

simulations produce each of the histograms. Current and synaptic activations begin at TTS=0. Notice the scale difference.

**Thermal Noise**

Thermal noise is also present, but it is the noise source of least concern. Although thermal noise increases with increasing resistance, the low-pass property of a resistive-capacitive circuit when recording across the capacitor exactly cancels out the increase in thermal noise by lowering the high-frequency cutoff of the filter [39 p352]. Thus, the thermal noise (calculated as the expected value of the variance of the voltage) is equal to $kT/C = 1.78 \times 10^{-11}$ $V^2$ (C is the capacitance of the neuron, 240 pF; $k$ is the Boltzmann constant; and $T$ is body temperature, 310 K). Thus, the standard deviation of this zero-centered noise is 4.22 µV. Compare this value to the shot noise fluctuations shown in Fig 5 inset; it is much smaller than the sodium channel shot-noise.

**I(Λ;TTS)**

Treating the neuron as an information channel, we identify the input random variable as $\Lambda$ and the output random variable as TTS. A proper calculation requires an appropriate range for both of these random variables. As in the fit to the histogram (Fig 4), the assumed range of TTS is the positive real line; however, as one can see in Fig 4, the probability of long and short durations is miniscule. Even though the earlier results demonstrate an expanded range for the intensities of synaptic activation beyond those reported by Hu et al, this range seems overly modest for a pyramidal neuron of neocortex whose average activity is under 10 Hz, implying an average firing time between pulses of more than 100 ms. Regardless of the biophysical failings of the model, an appropriate approximation for a meaningful mutual information requires us to extend the range of synaptic intensities beyond those values appearing in Fig 1. Therefore, we extended the range on the low-intensity end to 200 ms. The effect this has on the mutual information calculation is shown in S3 Fig.

Since we do not know the probability distribution for Λ, we compare the result for three different distributions. The distributions and each of their associated mutual information values are summarized in Table 1. The calculation is described in detail in the methods section.

**Table 1 Distributions for Λ and associated mutual information values.**

| Distributional Form | Distribution | E[Λ] (events/ms) | H(T) (bits/spike) | Mutual Information (bits/spike) |
|---|---|---|---|---|
| $c/\lambda$ | $1.07/\lambda$ | 54.2 | 5.88 | 3.00 |
| $c$ | 0.020 | 58.1 | 5.37 | 2.77 |
| $\dfrac{Exp(-\lambda/c)}{c \cdot [Exp(-\lambda_{min}/c) - Exp(-\lambda_{max}/c)]}$ | $\dfrac{Exp(-\lambda/100)}{28.59}$ | 55.9 | 5.94 | 2.99 |

Common to all distributions is the range of Λ: $\lambda \in [32.78, 83.33]$.

## Discussion

**Linearity**

The fitting of the inverse Gaussian distribution and linearization is important in the historical context of neuron modeling. It is a pervasive assumption in many computational models, dating back to McCulloch and Pitts [1], that linear computation (or log-linear) in fact, is assumed and seemingly required for several recent models [40-42]. As a first qualification of linear integration and interspike interval (ISI) coding, the inverse of the TTS should be linearly related to the intensity of the current-step. Previously, an earlier model of a simplified neuron model used voltage-gated conductances to demonstrate linear charging due to the added voltage-dependent nonlinearities [17]. The idea of a neuron with nonlinear, voltage-activated conductances is at odds with the textbook RC neuron whose synaptic excitation is interpreted as an Ornstein-Uhlenbeck process [7-12,43]. Until proved otherwise, it seems safe to assume that all excitatory neocortical neurons contain voltage-activated conductances.

The results show that over a limited range of intensities this model produces the desired inverse relationship between intensity and TTS (Fig 1) in a consensus pyramidal model, and these observations were made without manipulating the model's values and placements of voltage-gated conductances. Additionally, the model produces an inverse Gaussian distribution for the TTS (Fig 4). Indeed, this idea of the inverse Gaussian and the linearity required of the excitation process is actually quite old (a result first pointed out in neuroscience by Gerstein and Mandelbrot) [2]. This result is available to an Ornstein-Uhlenbeck diffusion, but only as an approximation of a diffusion that runs for a very short time [44]. In sum, the observations here are a prime example of complexity (the voltage-activated conductances) leading to simplicity (linear additivity producing a closed-form probability hitting time distribution).

**Sources of Noise**

Some recent reviews have speculated on various qualitative sources of noise [43-46]. Here we are concerned with a more precise, quantitative statement than these qualitative speculations. Historically, perhaps the earliest measurement in estimation of noise sources is found in [29], who suggest that synaptic noise can account for all the variability in ISIs. Because their observations were performed *in vivo*, their results cannot directly isolate the noise contributions from the synaptic input and channel noise; however, the observations here are able to do so (Fig 3). Even more recent results to do not contradict the idea that synaptic noise dominates over channel-noise sources [49]. Although their results suggest non-Poisson inputs, these results and the interpretations are consistent with synaptic variability dominating VAC shot-noise variability in the physiological situation. In that study, the input excitation of the neuron creates a near steady-state situation (current applications last for at least a second). These long-duration stimuli will produce very different effects than our very short current applications that are of most relevance for studying TTS.

Under a very different set of assumptions, Sengupta et al. [50] also conclude that VAC shot-noise is only a minor and possibly ignorable source of input-output variability. Two distinctions stand out when comparing their results to the current study. Their model neuron is a biophysical neuron but relative to the one studied here seems somewhat arbitrary in its construction. The neuron used here is the consensus layer 5 pyramidal neuron (see third paragraph of introduction for references). Second, they model a dynamic input signal, perhaps inspired by the fly eye research that used a Gaussian input signal and symbolic decoding [34,51]. In contrast, the neuron of interest here can be thought of as performing computation (perhaps discrimination) over the interval of a visual fixation. In this case, we must suppose a prior distribution. Because we don't know the correct distribution, we offer three possibilities each of which is compatible with the dynamic range of the neuron being studied. Of course, any calculation of mutual information must make an assumption about the long-term prior distribution of the input intensity ($\Lambda$) for the neuron under study.

While it is certainly true that noise limits the detection by peripheral sensors (hair cells in the basal ganglia, cones in the eye, etc.), we speculate that randomization processes rather than classic versions of noise itself are of much more concern in calculating information transmitted between neurons than explicit noise processes including thermal noise and shot noise of voltage-activated conductances. Other types of noise might include chaotic oscillations generated by interactions between voltage-activated conductances, but we assume in pyramidal cell neocortical computation there are no such oscillations. Thus, rather than calling the unpredictable firing of a neuron noise, we might use the term randomization, which is produced by the large number of poorly synchronized, input-line activations. Moreover, any such randomization process which itself is enhanced by quantal synaptic failure [52].

**Mutual Information**

Our estimate of mutual information agrees with the approximately 3 bits per spike calculated by [34]. Indeed, this agreement might be somewhat surprising considering how large a dynamic range they used versus how restricted the dynamic range is for the pyramidal neuron studied here. Other estimates range from 2 – 5 bits per spike [53-60]; see [61] for a review.

It can be argued that the 3-bit calculation here is an underestimate due to the constraints imposed on the consensus neuron being used. For example, the model considered here sustains firing only over a limited range of input intensities, and a more accurate neuron with more voltage-activated conductances (e.g. the $I_h$ conductance or even a large set of spatially-distributed conductances [62]) might lead to a neuron that has a greater dynamic range. Another limitation of the current study is a lack of inhibition. Physiologically, a neocortical neuron will simultaneously receive inhibition and excitation. Such inhibition downgrades the effect of synaptic events, which in turn requires larger values of $\Lambda$ to produce the same firing rates ($\Lambda$). By increasing the number of events per output spike, the information rate per output spike goes up. For example, a four-fold increase in the number of events needed to fire the neuron increases the mutual information by about 1 bit / spike.

Other assumptions affect the calculation of mutual information. Stochastic $K^+$ channels could contribute to the variance of the TTS, thus decreasing mutual information. However, the TTS reflects variability occurring at low amounts of depolarization compared to when voltage-dependent $K^+$ channels tend to be activated. Additionally, $K^+$ channels can have smaller single-channel conductances than the $Na^+$ channels used here [63-65], and thus their noise contribution will be smaller.

**Suggestions for future research**

This research invites more research into the linear model of the neuron. As a relatively simple alternative to the stochastic, leaky-neuron model, the linear model can provide insights into neural computation, even in up-to-date, realistic physiological models. Before going further with this model neuron, experimental scientists should evaluate the spike-generating characteristics of this layer V neuron over a wider range of conditions. These conditions include both higher and lower levels of excitation, excitation in the presence of inhibition, and TTS relationships after a preceding spike.

## Methods
### Model

The cell that is modeled is a Layer 5 pyramidal cell quantified by Mainen & Sejnowski [20] (their Fig 1D). The biophysical model evolved from [65] in which it was used to study back propagating action potential spikes. This led to [19] in which it was further refined (model can be retrieved from ModelDB [66]). The model was fit to measurements at several different places on the neuron including the AIS, axon, and dendrites.

The model includes both Nav 1.6 and Nav 1.2 channels distributed along the AIS. It is well-accepted that the action potential is initiated at the distal AIS [26-27,67] with its high concentration of different sodium channel types [25,28,68]. The capacitance in the soma was changed from 1 $\mu F/cm^2$ to .9 1 $\mu F/cm^2$, as this is a more realistic value [69]. In addition, a small amount of persistent Na channels was added at the proximal apical dendrite to simulate more realistic neuronal firing over a longer range of times. The persistent sodium channels were modelled using the Nav1.6 channels with shifted inactivation rate equations, so that they would inactivate much more slowly [38,67,70].

The main change was to alter the sodium channels so that they operated stochastically. The stochastic sodium channels were modeled with a widely-used eight-state kinetic reaction scheme describing the $m^3h$ Hodgkin-Huxley activation kinetics [71-72] quantified in [73]. This gating scheme is shown in [74] (their Fig 4A). When run deterministically, there is no variability. These changes did not substantially change the overall shape of the action potential (Fig 5). The main difference is that the stochastic voltage-trace lies below the deterministic immediately before firing. Since it is noisier, the stochastic neuron can suddenly cross threshold with a stochastic event, and thus does not have to gradually increase to threshold as the deterministic neuron does. Thus, the stochastic neuron tends to be at a slightly lower voltage when it fires than the deterministic neuron.

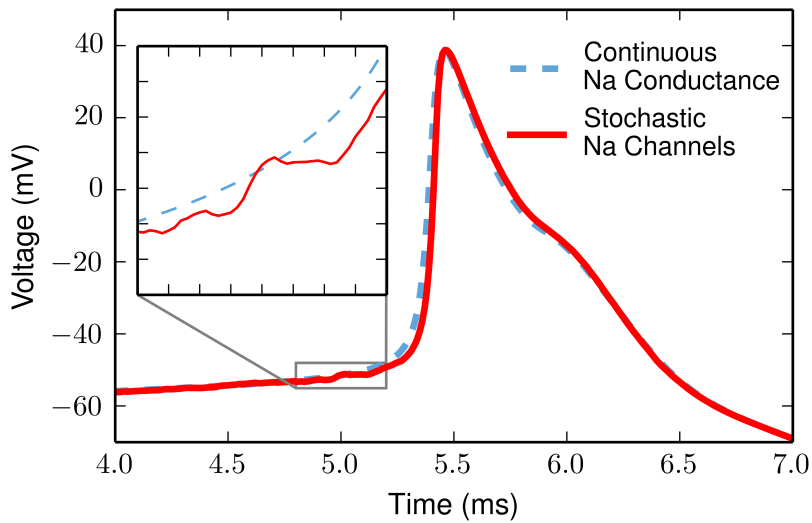

**Fig 5. Comparison of a stochastic- and a deterministic-based action potential.** The deterministic action potential (blue dashed line) reproduces the result of Hu et al; their action potential initiates at the AIS and spreads to the soma and apical dendrite. Aligned, peaked to peak, is a second action potential (solid red line) using stochastic Na-channels (both Nav 1.2 and Nav 1.6). Both action potentials are generated by the same somatic current-step of 1 nA. Inset y-axis goes from -55 mV to -48 mV (increments of 1 mV); inset x-axis goes from 4.8 ms to 5.2 ms (increments of .05 ms).

**Excitation**

Simulations were performed using the simulation environment *NEURON* [75]. A time step of 1 μs was used, and the results did not change for smaller time steps. In all cases, the neuron is allowed to come to steady state before the stimulus is applied. The neuron was stimulated in the main apical dendrite 250 μm from the soma because this is where synapses are located, and our aim was to study the noise in the synaptic input. This was done in two ways: first with a noise-free current-step in the main apical dendrite, and second by simulating synaptic activity. Synaptic activity was simulated using synapses distributed evenly along the main apical dendrite 200 μm-300 μm from the soma. Synapses were simulated as the Poisson arrival of square pulses of 200 pS that lasted 1.2 ms with a reversal potential of 0 mV. The Poisson assumption used here arises not from a Poisson assumption on individual inputs, but the Poisson approximation [76] produced by the unioning of all the input lines, each one being a point process.

As a surrogate for interpulse interval, the time-to-spike (TTS) is measured for current-step and for step random synaptic activations. The TTS is defined as the time it takes to achieve maximum voltage from steady state. This is used because we require a clean system for quantifying the noise effects of the sodium channel and the synaptic input fluctuations.

**Mutual Information Calculation**

Treating the neuron as an information channel, we identify the input random variable as $\Lambda$ and the output random variable as TTS. In order to calculate the mutual information $I(\Lambda;TTS)$ we require three probability distributions: $P(\Lambda)$, $P(TTS|\Lambda)$, and $P(TTS)$.

**P($\Lambda$).** We choose the distribution of $P(\Lambda)$ over the range we are studying. Since we do not know the probability distribution for $\Lambda$, we try multiple different distributions.

**P(TTS|$\Lambda$)**. We have seen that at a given intensity $\lambda$, the distribution of TTS follows an inverse Gaussian distribution (Fig 2). The PDF of the inverse Gaussian distribution with a mean of $\mu$ and a shape parameter of $\rho$.

$$IG(\mu,\rho) = \left(\frac{\rho}{2\pi TTS^3}\right)^{1/2} \exp\left(\frac{-\rho(TTS-\mu)^2}{2\mu^2 TTS}\right) \quad (1)$$

The shape parameter $\rho = \mu^3/\sigma^2$, where $\sigma^2$ is the variance of TTS. To get an expression for $P(TTS|\Lambda)$, we require the parameters $\mu$ and $\rho$ for any $\lambda$ in our range. To find $\mu$ as a function of $\lambda$ ($\mu(\lambda)$), we do an inverse best-fit to the the collected data on $\lambda$ vs. $E[TTS]$, and to find $\sigma^2$ as a function ($\sigma^2(\lambda)$), we do an inverse best-fit to the collected data on $\lambda$ vs. $Var[TTS]$. If we need to extrapolate the variance past the range of collected data, we use the largest observed variance (18.57 ms$^2$). All fitting data is in S1 Table. Then we can define $P(TTS|\Lambda)$.

$$P(TTS|\Lambda) = IG\left(\mu(\lambda), \frac{\mu(\lambda)^3}{\sigma(\lambda)^2}\right) \quad (2)$$

**P(TTS).** $P(TTS)$ is computed by numerical integration over the entire range of $\Lambda$ under consideration. The resulting $P(TTS)$ distributions are shown in S4 Fig.

$$P(TTS) = \int_{\lambda_{min}}^{\lambda_{max}} P(TTS|\Lambda=\lambda) \cdot P_\Lambda(\lambda) d\lambda \quad (3)$$

**I($\Lambda$;TTS).** The mutual information (in bits / spike) is calculated according to the following equation by summing over $\Lambda$ with steps of .1 events/ms and by summing over TTS in the range [1 ms, 250 ms] with steps of 0.05 ms.

$$I(\Lambda;TTS) = \int_{32.78}^{83.33} P_\Lambda(\lambda) \int_0^\infty P(TTS|\Lambda) \log_2 \frac{P(TTS|\Lambda=\lambda)}{P(TTS)} dTTS d\lambda \quad (4)$$

**Acknowledgements**

We thank Costa Colbert for helpful suggestions and comments during the research and the writing of the manuscript. We also thank the Department of Neurosurgery for their support. This work was supported by the National Science Foundation 1162449 – Toby Berger & William Levy.

# Supporting Information
**S1 Table. Values for step intensity and recorded statistics on TTS**

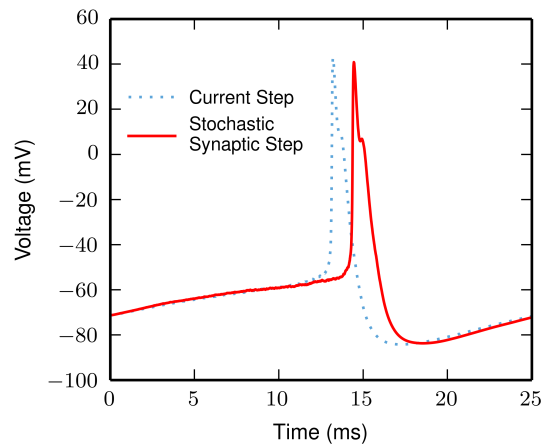

**S2 Fig. Voltage traces for a current-step and stochastic synaptic-step.**
Voltage is measured at the AIS. Current-step is .7 nA and stochastic synaptic step is 58.33 events / ms. Though the average input current is the same, this stochastic step generally fires after this current step (see histograms in Fig 5) This delay occurs because the synaptic activations are distributed over a 100 μm section of the dendrite while the current-step is a point source in the dendrite.

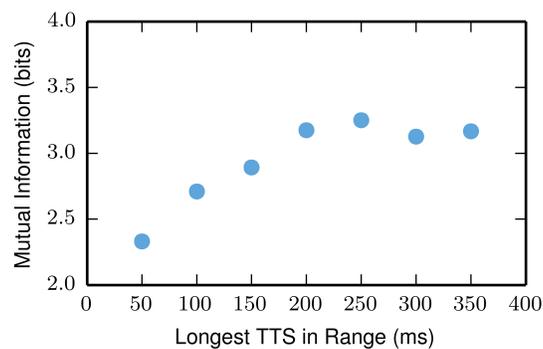

**S3 Fig. Mutual information values over different extrapolated ranges.**
All calculations share the same shortest TTS (6.8 ms) and use the same distributional form for $\Lambda$ ($c/\lambda$, where c is a constant). The maximum value of $\lambda$ is 83.33 events / ms. The minimum value of $\lambda$ changes depending upon the longest TTS in the range, and the value of c is chosen to normalize the distribution.

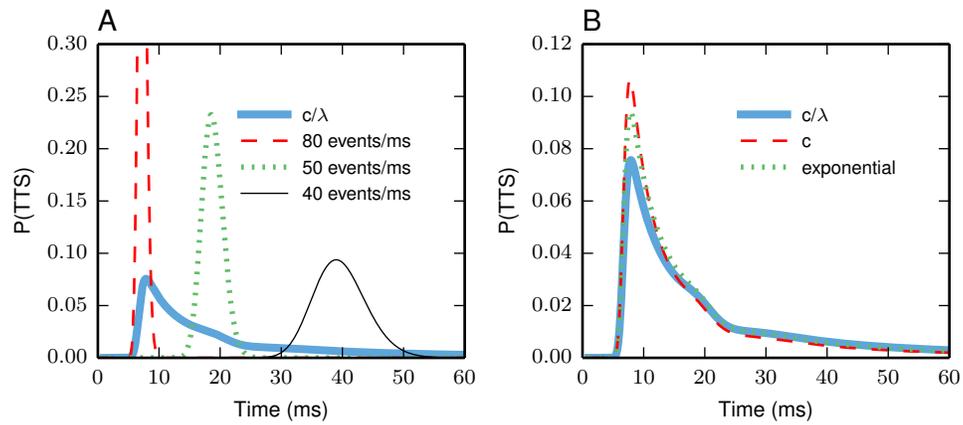

**S4 Fig. P(T) distributions**. (A) The P(TTS) distribution (thick blue curve) is computed as the weighted sum of conditional distributions. Three conditional distributions P(TTS|λ) are shown for λ=80 events/ms (red dashed line), λ=50 events/ms (green dotted line), and λ=40 events/ms (thin black line). (B) P(TTS) resulting from different $\Lambda$ distributions. Each distribution for T corresponds to a distribution for λ in Table 1.